\documentclass[floatfix, superscriptaddress,preprint,aps]{revtex4}

\usepackage{siunitx}
\usepackage{graphicx}
\usepackage{color}
\usepackage{multirow}
\usepackage[normalem]{ulem}
\usepackage{todonotes}
\usepackage{array}

\begin{document}

\title{Photoluminescence enhancement by deterministically site-controlled, vertically stacked SiGe quantum dots}

	\author{Jeffrey Schuster$^\ddagger$}
	\author{Johannes Aberl}
	\author{Lada Vuku\v{s}i\'{c}}
	\author{Lukas Spindlberger}	
	\affiliation{Institute of Semiconductor and Solid State Physics, Johannes Kepler University Linz, Altenbergerstra{\ss}e 69, 4040 Linz, Austria}
	\author{Heiko Groiss}
	\affiliation{Christian Doppler Laboratory for Nanoscale Phase Transformations, Center for Surface and Nanoanalytics (ZONA), Johannes Kepler University Linz, Altenbergerstra{\ss}e 69, 4040 Linz, Austria}
	\author{Thomas Fromherz}
	\author{Moritz Brehm}
	\author{Friedrich Sch\"affler}
	\affiliation{Institute of Semiconductor and Solid State Physics, Johannes Kepler University Linz, Altenbergerstra{\ss}e 69, 4040 Linz, Austria}

\begin{abstract}
The Si/SiGe heterosystem would be ideally suited for the realization of complementary metal-oxide-semiconductor (CMOS)-compatible integrated light sources, but the indirect band gap, exacerbated by a type-II band offset, makes it challenging to achieve efficient light emission. We address this problem by strain engineering in ordered arrays of vertically close-stacked SiGe quantum dot (QD) pairs. The strain induced by the respective lower QD creates a preferential nucleation site for the upper one and strains the upper QD as well as the Si cap above it. Electrons are confined in the strain pockets in the Si cap, which leads to an enhanced wave function overlap with the heavy holes near the upper QD's apex. With a thickness of the Si spacer between the stacked QDs below \SI{5}{\nano\metre}, we separated the functions of the two QDs: The role of the lower one is that of a pure stressor, whereas only the upper QD facilitates radiative recombination of QD-bound excitons. We report on the design and strain engineering of the QD pairs via strain-dependent Schr\"odinger-Poisson simulations, their implementation by molecular beam epitaxy, and a comprehensive study of their structural and optical properties in comparison with those of single-layer SiGe QD arrays. We find that the double QD arrangement shifts the thermal quenching of the photoluminescence signal at higher temperatures. Moreover, detrimental light emission from the QD-related wetting layers is suppressed in the double-QD configuration.

\end{abstract}
\maketitle

\section{Introduction}
The ever-increasing demand for data processing and transfer, as well as for on-chip sensing applications, have triggered far-reaching interests in the development of monolithically integrated optoelectronic platforms based on group-IV complementary metal-oxide-semiconductor (CMOS) technologies \cite{rickman_CommercializationSilicon_2014}. Today, silicon photonics provides building blocks for various optoelectronic applications and devices such as modulators, detectors, diodes, waveguides, and other essential functions demanded by present and future classical- and quantum technology \cite{siew_ReviewSilicon_2021, tsung-yangliow_SiliconModulators_2010, rauter_RoomTemperatureGroupIV_2018, juan-colas_ElectrophotonicSilicon_2016, aberl_SiGeQuantum_2019, vines_HighPerformance_2019, fischer_GrowthCharacterization_2015}. The main open challenge of silicon photonics \cite{pavesi_SiliconPhotonics_2004, lipson_GuidingModulating_2005, moss_NewCMOScompatible_2013,  rickman_CommercializationSilicon_2014} is the implementation of an efficient light source that operates at room temperature, can be monolithically integrated into Si devices with standard CMOS technology and exhibits emission wavelengths in the near infra-red spectral range. In particular, such a light source would allow the implementation of on-chip optical interconnects to overcome dissipation losses and the limited bandwidth of electronic signal transmission caused by on-chip copper wiring \cite{assefa_CMOSIntegrated_2011}. Although hybrid solutions involving III-V materials bonded to silicon-on-insulator (SOI) platforms have been reported \cite{chen_ElectricallyPumped_2016, wang_RoomtemperatureInP_2015}, most of the advantages associated with monolithic integration are lost, in particular overlay accuracy, material and fabrication cost, as well as yield. Covering the optical telecommunication bands between \SI{1.3}{\micro\metre} and \SI{1.6}{\micro\metre} \cite{tsybeskov_SiliconGermaniumNanostructures_2009, hauke_ThreedimensionalSilicon_2012, schatzl_EnhancedTelecom_2017}, the CMOS-compatible Si/SiGe heterosystem is, in principle, well-suited for optoelectronic on-chip integration. However, both the indirect bandgap and the type-II band offset of this heterostructure \cite{abstreiter_StrainInducedTwoDimensional_1985} render it challenging to achieve high quantum efficiencies. In SiGe quantum dots (QDs) \cite{sunamura_GrowthMode_1995, schittenhelm_PhotoluminescenceStudy_1995} the radiative recombination rate is enhanced by the low-dimensional carrier confinement, which increases the local density of states and relaxes the wave-vector-conservation criteria \cite{yu_FundamentalsSemiconductors_2010}. Over the last two decades, epitaxial growth techniques have been developed to tailor the essential properties of SiGe QDs, such as their size, shape, and composition \cite{schmidt_EffectOvergrowth_2000, stangl_EffectOvergrowth_2003, montalenti_AtomicScalePathway_2004, stoffel_ShapeFacet_2004, brehm_KeyRole_2009}. Moreover, it has also become possible to control the exact position of the SiGe QDs on a Si substrate by providing lithographically defined nucleation sites \cite{zhong_TwodimensionalPeriodic_2003, zhang_SiGeGrowth_2007, brehm_PhotoluminescenceInvestigation_2015, brehm_SitecontrolledAdvanced_2017, smagina_NucleationSites_2018, novikov_OneStageFormation_2021}.

While positioning with nanometer accuracy is a precondition for monolithic CMOS integration, it also allows exploitation of the Purcell effect \cite{purcell_ResonanceAbsorption_1946} via deterministic coupling of SiGe QDs to photonic crystal resonators \cite{schatzl_EnhancedTelecom_2017, zeng_SingleGermanium_2015, smagina_LuminescenceSpatially_2020}. Still, the type-II band offset causes carrier separation in real space, with holes localized in the SiGe regions and weak electron confinement in the surrounding Si matrix \cite{tsybeskov_SiliconGermaniumNanostructures_2009, penn_ApplicationNumerical_1999}. Compared to a type-I heterosystem, the overlap of the wave functions is significantly reduced, and thus is the matrix element for radiative recombination \cite{yu_FundamentalsSemiconductors_2010}. A recent approach to address this problem utilizes deep-level Ge-ion-implantation into SiGe QDs, facilitating quasi-direct optical transitions \cite{brehm_SitecontrolledAdvanced_2017, grydlik_LasingGlassy_2016, grydlik_LaserLevel_2016, groiss_PhotoluminescenceEnhancement_2017, rauter_RoomTemperatureGroupIV_2018, spindlberger_InSituAnnealing_2020, spindlberger_AdvancedHydrogenation_2021}. This approach has been highly successful in achieving pronounced room-temperature light emission with further improvements achieved by annealing \cite{spindlberger_ThermalStability_2019} and defect passivation in the surrounding Si matrix \cite{spindlberger_InSituAnnealing_2020, spindlberger_AdvancedHydrogenation_2021}. However, deep-level implantation has so far only been demonstrated with the hut-cluster-type \cite{mo_KineticPathway_1990} of SiGe QDs which have to be grown at temperatures too low to allow for efficient site-control \cite{brehm_SitecontrolledAdvanced_2017}. 

To preserve the option of perfect site-control, we propose in this contribution a different path toward enhanced optical transition matrix elements and improved temperature stability of SiGe QDs: We engineer the built-in strain caused by the different lattice constants of Si and Ge \cite{abstreiter_StrainInducedTwoDimensional_1985, schaffler_HighmobilitySi_1997} by introducing a layer sequence with vertically close-stacked SiGe QD pairs \cite{schmidt_MultipleLayers_2000, teichert_StressinducedSelforganization_1996, zhang_StrainEngineering_2010, zinovieva_ElectronSpatial_2019}. In this way, we aim at increased wavefunction overlap of the confined electrons and holes, and thus at enhanced radiative recombination. 

This approach becomes feasible because a thin Si layer grown above a SiGe QD array on a strain-free Si (001) substrate experiences tensile strain in the (100) growth plane with local maxima above the apices of the QDs \cite{schmidt_MultipleLayers_2000, zhang_StrainEngineering_2010}. Deposition of a second Ge layer above this locally strained Si film will then lead to the nucleation of SiGe QDs at the local strain maxima \cite{schmidt_MultipleLayers_2000, zhang_StrainEngineering_2010}. As a result, the two QD layers become vertically stacked with high overlay accuracy and separated by a well-defined Si spacer layer \cite{schmidt_MultipleLayers_2000, zhang_StrainEngineering_2010}.

Locally, the upper QDs induce increased tensile strain in a subsequently deposited Si capping layer that is further enhanced by Ge accumulation along the growth direction \cite{schmidt_MultipleLayers_2000, zhang_StrainEngineering_2010}. The tensile in-plane strain in the Si cap splits the sixfold degenerate Si conduction band into the four energetically raised in-plane valleys ($\Delta_{xy}$), and the two energetically lowered $\Delta_{z}$ valleys along the growth direction \cite{abstreiter_StrainInducedTwoDimensional_1985, schaffler_HighmobilitySi_1997}. The goal here is to separate the functionalities of the two QDs in the stack: The lower SiGe QD acts solely as a stressor for the upper QD, where the holes become confined, as well as for the adjacent Si cap layer, where the $\Delta_{z}$-electrons are localized. Radiative exciton recombination occurs between these confined carriers.

In the following, we report on the simulation-assisted sample design, the realization of such layer sequences by solid-source molecular beam epitaxy (MBE) on pre-patterned Si substrates, as well as their characterization by atomic force microscopy (AFM), scanning transmission electron microscopy (STEM), and composition-mapping by selective wet chemical etching (nanotomography). The main emphasis of this study is put on a comparative assessment of single- and double QD samples by position- and temperature-dependent micro-photoluminescence (\textmu-PL) experiments.

\section{Sample Design}

The commercial Schr\"odinger-Poisson-solver nextnano++ \cite{birner_NextnanoGeneral_2007} with a single-band model was employed to design layer sequences for subsequent implementation by epitaxial growth. Details regarding the simulations performed in this work are given in the "Methods" and "Simulations" sections.

For the double-QD configuration, we performed an initial series of simulations with varying Si spacer thickness to assess the role of the strain pocket in the Si spacer layer above the lower QD. Generally, we found that the Si spacer has to be less than \SI{5}{\nano\metre} thick to suppress detrimental $\Delta_z$-electron confinement there. We restricted our experimental investigations to this range of thin spacer layers to make sure that only the upper QD in the stack is optically active. This condition is certainly no longer fulfilled for spacer thicknesses exceeding \SI{10}{\nano\metre}, for which distinguishable excitonic transitions from both QDs in the stack have been reported \cite{schmidt_MultipleLayers_2000}. Also, the QDs should be as small as possible to achieve zero-dimensional carrier confinement and enhanced wavefunction overlap.

\section{Sample Fabrication}

Based on the initial simulations, a series of six samples was grown by MBE on Si(001) samples that were pre-patterned by electron beam lithography and reactive ion etching (Table 1). Details of the fabrication procedure are given in the "Methods" section.

In brief, each sample contained a large number of simultaneously fabricated pit arrays, with the pit radius and period being systematically varied from array to array. The geometry and the growth parameters were chosen such that a single QD or a stacked QD pair (Table \ref{tab:samples}) nucleates in the center of each pit during subsequent overgrowth by MBE. The vertical separation of the QD stacks was varied by introducing a Si spacer with a nominal thickness of 2, 4 and \SI{5}{\nano\metre}, respectively. To achieve small QDs, the amount of deposited Ge for each QD layer was chosen to be just above the minimum amount for QD formation, i.e., \SI{5.6}{\angstrom} of pure Ge for the first, and, to account for Ge segregation, \SI{4.2}{\angstrom} Ge for the second QD layer \cite{groiss_PhotoluminescenceEnhancement_2017, schmidt_MultipleLayers_2000}. Also, a relatively high temperature of \SI{670}{\celsius} was chosen for QD growth to minimize the formation of non-radiative recombination centers. Samples A-D were capped with Si for characterization by STEM and \textmu-PL. Samples RefA and RefB remained un-capped to provide direct access to the QDs on the surface via AFM and nanotomography experiments.

\begin{table}
\begin{center}
\caption{\label{tab:samples}Upright layer sequences of the samples used in this study. Samples A and RefA incorporate a single QD layer only, all other have vertically stacked QD pairs separated by the denoted Si spacer thicknesses. Samples RefA and RefB have no Si cap to allow for AFM imaging and nanotomography. All other samples are Si-capped for PL experiments and STEM imaging. The growth parameters for the individual layers are given in the "Methods" part of the text.}
\begin{tabular}{ l|c|c|c|c|c|c } 
 &\multicolumn{6}{c}{sample \#} \\
 \hline
 layer & A & B & C & D & RefA & RefB \\
 \hline
 Si cap & yes & yes & yes & yes &  no & no \\ 
 QD 2 & no & yes & yes & yes & no & yes \\ 
 Si spacer & no & \SI{2}{\nano\metre} & \SI{4}{\nano\metre} & \SI{5}{\nano\metre} & no & \SI{2}{\nano\metre} \\ 
 QD 1 & yes & yes & yes & yes & yes & yes \\ 
  \multicolumn{7}{l}{Si substrate and Si buffer layer} \\
 \hline
\end{tabular}
\end{center}
\end{table}

\section{Structural Characterization}

\begin{figure}
\begin{center}
	  \includegraphics[width=\linewidth]{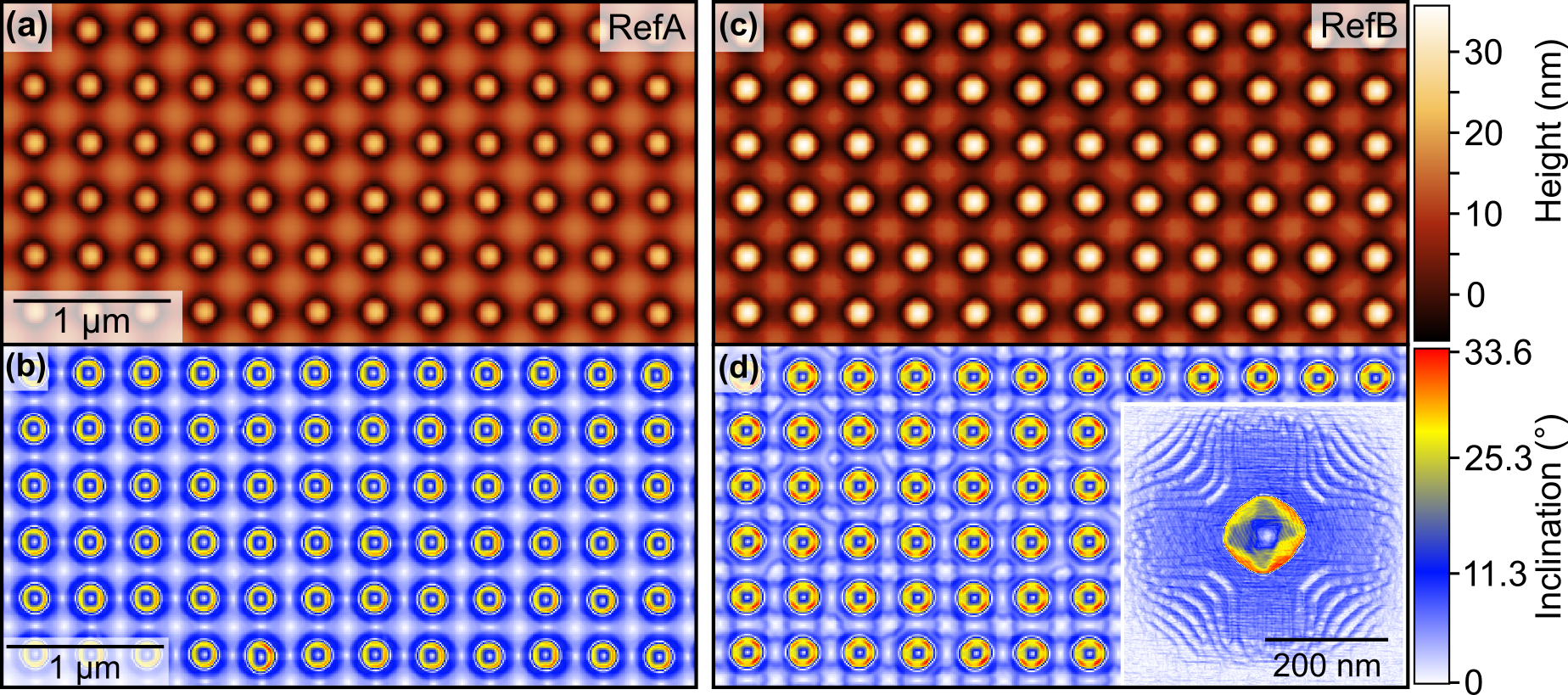}
  \caption{(a), (b) AFM micrographs of the un-capped samples RefA (left) and RefB (right), respectively, from the pre-patterned areas with a pit period of \SI{340}{\nano\metre}. The inclination maps (c) and (d) show the characteristic facets of dome-shaped dots, i.e., \{15~3~23\} indicated in red, \{1~1~3\} shown in yellow and the shallow top with a \{1~0~5\} facet (blue). Inset: Enlarged image of one pit with its dome-shaped QD showing the facets in higher magnification as well as the decoration of the pit sidewalls with \{1~0~5\}-facetted ripples (blue) and (001) terraces (white).}
  \label{fig:afm}
\end{center}
\end{figure}

Figure \ref{fig:afm}(a) and (c) show AFM micrographs of the uncapped reference samples RefA (single QD layer) and RefB (QD stack with \SI{2}{\nano\metre} spacer) from pre-patterned areas with a period of \SI{340}{\nano\metre}. In either case, perfect ordering in registry with the etch-pits in the substrate has been achieved. Figure \ref{fig:afm}(b) and (d) show corresponding inclination-angle maps of the QD arrays and an enlarged unit cell of sample RefB in the inset of Fig. \ref{fig:afm}(d).

\begin{figure*}
\begin{center}
  \includegraphics[width=\linewidth]{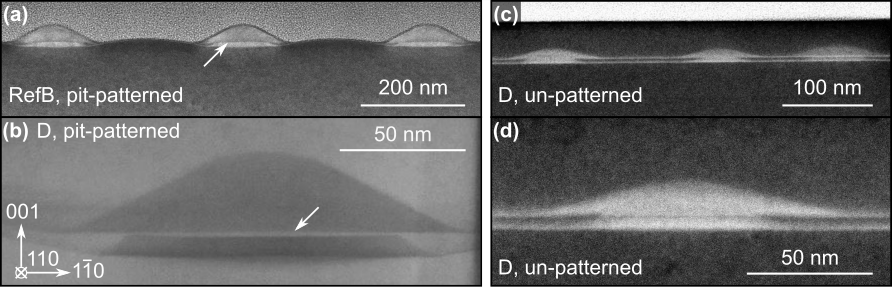}
  \caption{(a) HAADF image of ordered QD stacks in pits. (b) Bright-field image of a QD-stack in a pit. (c) HAADF image of randomly nucleated QD-stacks from an un-patterned area of Sample D. (d) Zoom-in of a QD-stack of image (c).}
  \label{fig:TEM}
\end{center}
\end{figure*}

To access the cross-sectional geometry of the fabricated samples, we performed STEM and high angle annular dark field (HAADF) imaging (see "Methods" section for details). Figure \ref{fig:TEM}(a) shows an HAADF image of three adjacent QD pairs from the pit-patterned region of the un-capped sample RefB with a \SI{2}{\nano\metre} thick Si spacer. The lower QD (bright in the HAADF image) sits in the center of the shallow, \{1~0~5\}-facetted pit and is clearly separated by a thin, Si-rich layer (dark line in the HAADF image; marked by an arrow) from the upper QD. Figure \ref{fig:TEM}(b) shows a related image from one of the ordered QD-pairs of sample D with a \SI{5}{\nano\metre} thick Si spacer as a bright-field (BF) image. The thicker spacer enhances the separation of the dots (bright in the BF image; marked with an arrow). The images of both samples reveal that a large percentage of the lower QD has been flattened during growth into what appears to be a terminating (001) interface towards the Si spacer and the upper QD.

Figure \ref{fig:TEM}(c) shows a cross-sectional HAADF image from the un-patterned area of sample D with three randomly nucleated QD-pairs that vary much stronger in size than the QD-pairs in the pit-patterned area. As the investigated area is pit-free, the bottom of the lower QDs is flat, and both WLs are clearly resolved as flat, bright lines. A zoom-in of one of the QD-pairs of Fig. \ref{fig:TEM}(c) is depicted in Fig. \ref{fig:TEM}(d), again as a HAADF image. As in the pit-patterned areas, the lower QD developed a flat upper interface to the clearly resolved spacer layer and the upper QD. Because of the higher area-density of the randomly nucleated QD-pairs, their volume is correspondingly smaller than those of the ordered ones.

Our selection of very thin Si spacer layers in combination with Ge coverages just at the onset of QD nucleation, and the rather high deposition temperatures for the Ge layers, were motivated by samples optimized for excitonic radiation. The cross-sectional STEM images in Fig. \ref{fig:TEM} reveal that this particular combination of growth parameters results in a rather unique geometry of the QD-stack: The lower QD develops a flat interface to the clearly distinguished Si spacer and to the upper QD. Moreover, the upper QD has a significantly larger volume than the lower one although more Ge was deposited for the lower QD. Evidently, Ge segregation from the lower to the upper dot had occurred, as reported in the literature \cite{schmidt_MultipleLayers_2000}, but a very thin, Si-rich spacer remains between the QDs which develops into a terminating (001) facet.

We can rule out that the flattening of the lower QD has occured during the deposition of the Si spacer layer which was intentionally deposited at \SI{470}{\celsius} (see "Methods" section) to induce conformal overgrow \cite{brehm_InfluenceSi_2011}. We assume that the shape and size transformation of the lower QD occurs during deposition of the second QD at the relatively high temperature employed here. The Si caps of all samples were deposited below \SI{500}{\celsius} to preserve the dome shape of the single QD (sample A) and the respective upper QDs in samples B-D. Comparison of Figs \ref{fig:TEM}(a) (uncapped sample RefB) and Fig. \ref{fig:TEM}(b) (sample D) confirms that the shape of the capped QD is indeed conserved under these conditions. Experiments are presently under way to understand this kind of internal facet formation in the completely miscible Si-Ge alloy. The results are beyond the scope of this study and will be published elsewhere.

To determine the composition profiles of the QDs we used nanotomography experiments based on a combination of repeated wet chemical etching steps and AFM scans \cite{rastelli_ThreeDimensionalComposition_2008}. A Ge map can then be extracted by utilizing the composition-dependence of the etching rate in the employed etchant. For more details see "Methods" section and references therein.

\begin{figure}
\begin{center}
	  \includegraphics[width=\linewidth]{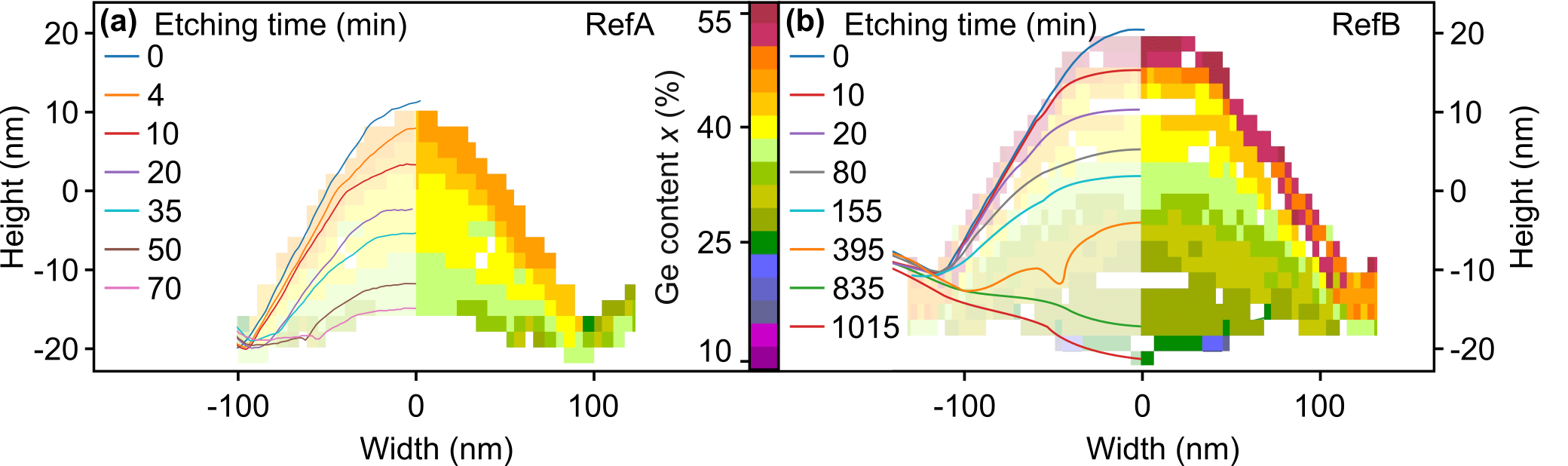}
  \caption{(a), (b) Nanotomography of samples RefA and RefB. The solid lines represent AFM line scans recorded after etching in NHH solution (see "Methods" section) for different etching times from \SIrange{0}{1015}{\minute}. The origin of the ordinate corresponds to the position of the (001) surface between the QDs. The underlying color-coded map shows the Ge-content which was extracted with the composition-dependent etching rate of the employed NHH solution. The negative half-space of the color map is a mirrored copy of the positive one.}
  \label{fig:nanotomo}
\end{center}
\end{figure}

Figure \ref{fig:nanotomo} shows a selection of AFM height profiles overlaying a color-coded composition map of QDs from the ordered regions of samples RefA (a) and RefB (b). The line profiles are labelled with the respective etching times in units of minutes. The extracted Ge-profiles (see "Methods" section) show increasing Ge-content in growth direction and from the center to the outside of the QDs. By comparing the results with STEM- and HAADF images, we assume that the Si spacer between the QDs in Fig. \ref{fig:nanotomo}(b) is located between the profiles recorded after 395 and \SI{835}{\minute}. The nanotomography experiments had to be performed on the un-capped samples RefA and RefB to get direct access to the QDs. However, the un-capped samples are expected to be terminated by about 2 monolayers (ML) of segregated Ge \cite{schmidt_MultipleLayers_2000} which leads to an overestimation of the topmost composition in the Ge map of Fig. \ref{fig:nanotomo} as compared to Si-capped samples.

\section{Simulations}

\begin{figure}
\begin{center}
  \includegraphics[width=\linewidth]{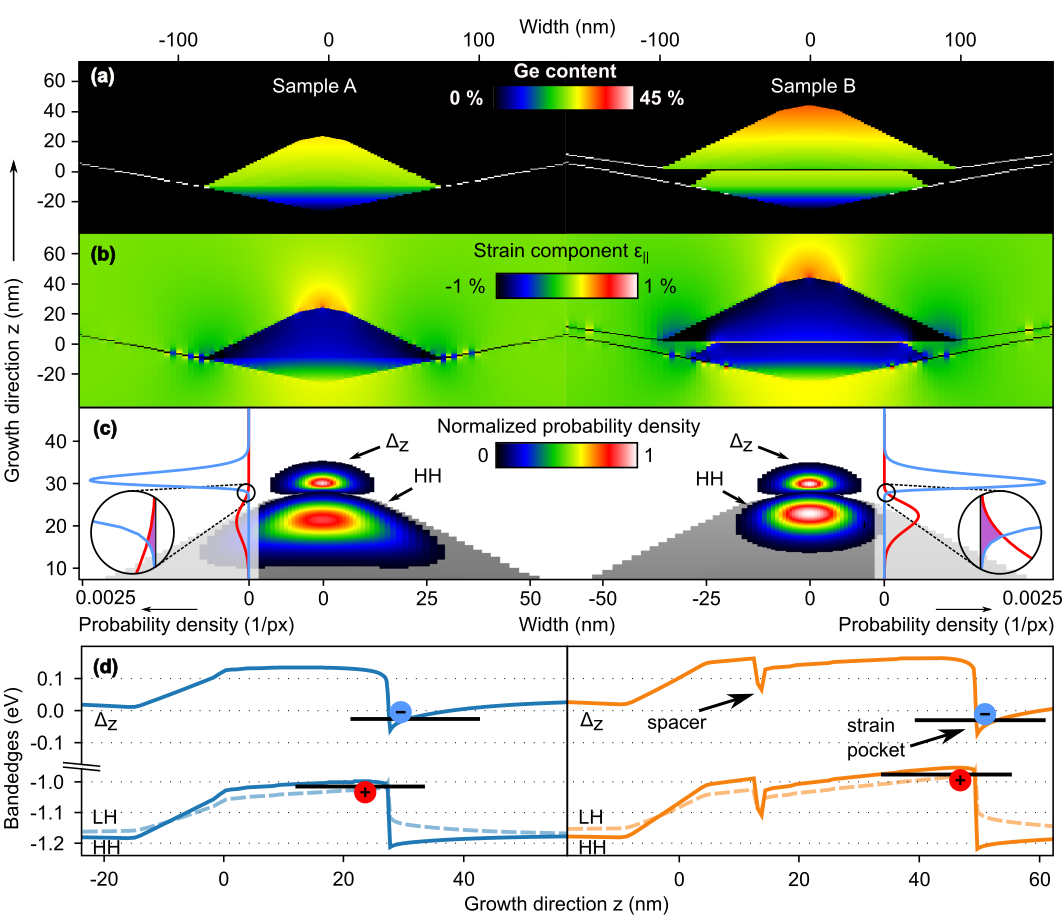}
  \caption{(a) – (c) Comparative nextnano++ simulations of a single Si$_{1-x}$Ge$_x$ QD (left) and a vertically stacked double QDs separated by a \SI{2}{\nano\metre} thick Si spacer (right) in a Si matrix. (a) Cross-sectional view of the layer stacks with color-coded Ge-content $x$. (b) Corresponding in-plane strain component $\varepsilon_{\parallel}$. The maximum tensile in-plane-strain in the Si matrix occurs above the Ge-rich apices of the QDs. (c) Normalized probability densities for heavy holes (HH) and $\Delta_z$-electrons for the QD sequences in (a) as color-code and line-plots. The HHs are located in the single and in the upper QDs, respectively, the $\Delta_z$-electrons above the respective apex. The color code is from black (vanishing probability) to white (maximum probability). The insets show the probability density per pixel, i.e., simulation node and an enhanced overlap for the double-dot structure which is highlighted by the inserted zoom-ins. (d) Schematic band edges for $\Delta_z$-electrons, HH and LH of both samples. The black lines indicate the ground state energy levels of $\Delta_z$-electrons and HH.}
  \label{fig:simulations}
\end{center}
\end{figure}

To assess the expected optical and electronic properties of the MBE-grown samples, we designed input models for nextnano++ simulations using a single-band model based on the experimentally determined geometries and composition profiles discussed above. The respective models for samples A and B are shown in Fig. \ref{fig:simulations}(a) as 2D projections of the 3D contour plots in a plane through the center of a single SiGe QD (left panel) and a closely stacked QD-pair with \SI{2}{\nano\metre} Si spacer (right panel). The color code represents the local Ge content which was modelled with lateral and vertical gradients based on the nanotomography results. The QDs are embedded into a matrix of pure Si ($x$ = 0) which is depicted in black in Fig. \ref{fig:simulations}(a). The lines at the bottom of each dot represent the Ge WLs that develop upon Ge deposition before the onset of QD nucleation  \cite{brehm_KeyRole_2009, brehm_QuantitativeDetermination_2008}. Their inclination represents the geometry of the \{1~0~5\}-facetted pit \cite{chen_InitialStage_2006} in which the QDs nucleate (Fig. \ref{fig:TEM}). 

The in-plane strain components $\varepsilon_{\parallel} = 1/2(\varepsilon_{xx} + \varepsilon_{yy})$ associated with the composition topologies in Fig. \ref{fig:simulations}(a) are depicted in Fig. \ref{fig:simulations}(b). Strain-free regions are displayed in light green, whereas the color-code from blue to red shades represents negative (compressive) and positive (tensile) strain values, respectively \cite{schaffler_HighmobilitySi_1997}. As expected, tensile strain regions develop in the Si layers above the apices of the QDs, with strain above the upper QD of the double-QD arrangement being significantly more pronounced.

To assess the optical matrix elements, the squares of the wave functions $\psi^\ast \psi$ (probability amplitudes) for the lowest energy states of $\Delta_z$-electrons and HH were calculated. The light holes (LH) can be neglected in the temperature range investigated in this study since their ground state is energetically unfavorable by 40 meV as compared to the HH ground state. Figure \ref{fig:simulations}(c) depicts the normalized probability densities with a color-code from black (zero) to white (maximum probability) and a lineplot of the probability densities per pixel (simulation node) along the center of the QDs. Because of the similarities of the single QD and the upper QD in the stack regarding the involved carriers, energy levels and geometries, the oscillator strengths will essentially be proportional to the overlap of the probability amplitudes of HH and $\Delta_z$-electrons \cite{yu_FundamentalsSemiconductors_2010}. The nextnano++ simulations show that the overlap of the probability amplitude is indeed enhanced by a factor of three in the case of the QD stack (see inserts in Fig. \ref{fig:simulations}(c)). The $\Delta_{xy}$-electrons are not shown here, because they are confined around the bottom of the upper dot and, according to our simulations, are energetically too unfavorable to contribute to the PL signals.

Figure \ref{fig:simulations}(d) shows schematically the variations of the conduction- and valence band edges through the centers of the QDs along the growth direction. The most favorable locations for the HH are situated in the single and upper SiGe dots, respectively. The dashed lines depict the valence band edge for the LH which follow in principle the HH band edge but is shifted by about \SI{30}{\milli\electronvolt} to lower energies. The tensile strain pockets in the Si regions above the apex of each QD favor the localization of $\Delta_z$-electrons. The respective ground state energy levels are depicted as black lines. The iteration scheme employed for the simulations (see "Methods" section) ensures that the  simulated $\Delta_z$/HH transition energies agree with the experimental PL spectra discussed in the next section.

\section{Photoluminescence: Measurements and Results}

\begin{figure}
\begin{center}
  \includegraphics[width=\linewidth]{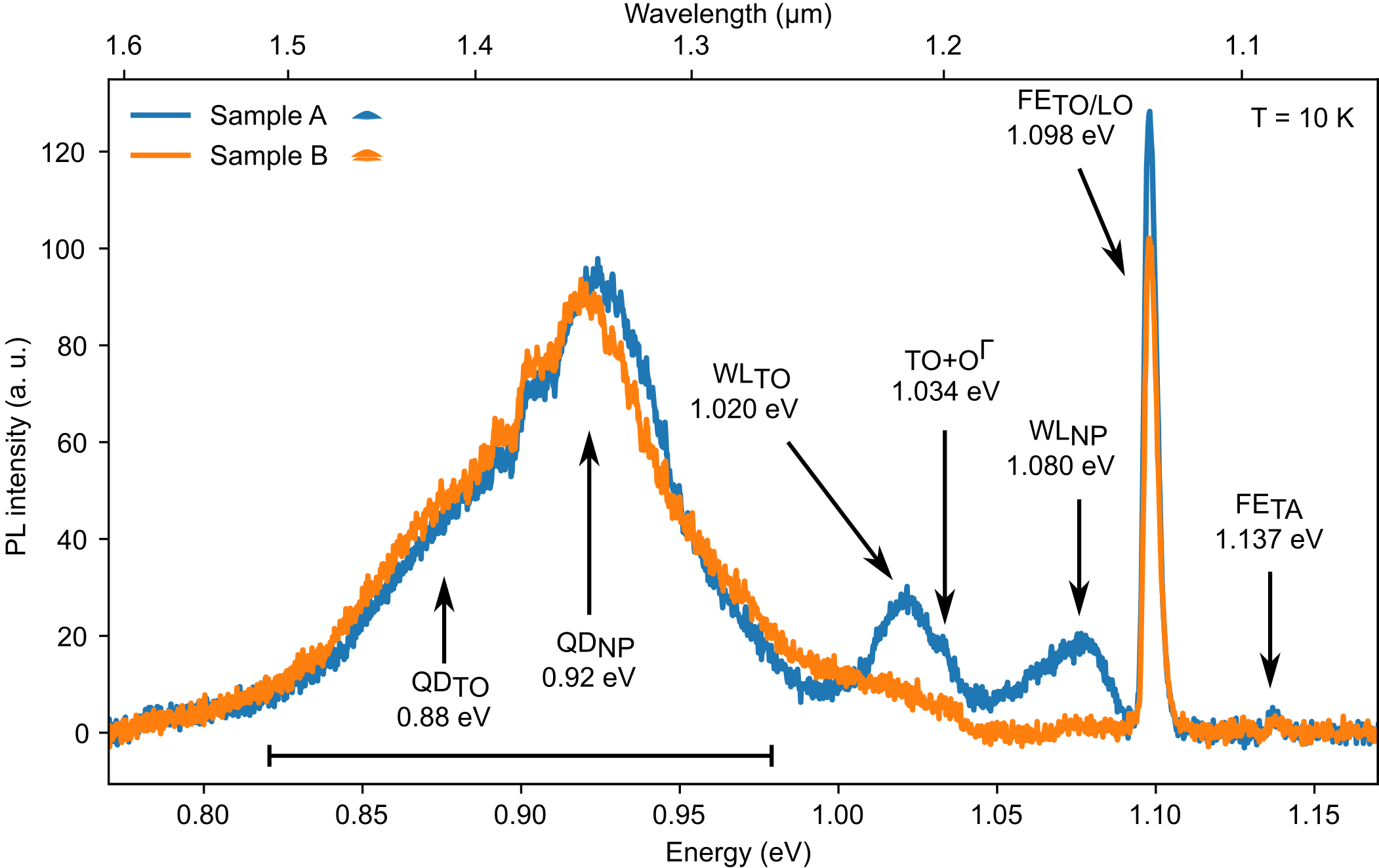}
  \caption{PL-Spectra from ordered QD arrays of samples A (blue) and B (orange) recorded at \SI{10}{\kelvin} with an excitation intensity of \SI{14.1}{\kilo\watt\per\square\centi\metre}. The signals from the QD arrays consist of a no-phonon signal (QD$_\mathrm{NP}$) at \SI{0.92}{\electronvolt} and a Ge-TO-phonon replica (QD$_\mathrm{TO}$) at \SI{0.88}{\electronvolt}. Sample A shows the no-phonon peak of the WL (WL$_\mathrm{NP}$) at 1.08 eV and its Si-TO-phonon replica (WL$_\mathrm{TO}$) at \SI{1.020}{\electronvolt}. All other peaks are substrate-related, as discussed in the text.}
  \label{fig:peaks}
\end{center}
\end{figure}

Position-, temperature- and excitation-intensity- (see Suppl. Information) dependent \textmu-PL experiments were performed on samples A-C (Table \ref{tab:samples}) in an experimental setup described in the "Methods" section. Figure \ref{fig:peaks} shows the \textmu-PL spectra of samples A and B which were recorded at \SI{10}{\kelvin} with an excitation-intensity of \SI{14.1}{\kilo\watt\per\centi\metre\squared}. Both spectra were obtained from the respective QD arrays with a period of \SI{340}{\nano\metre} and a pit radius of \SI{100}{\nano\metre}. The Si substrate contributes the optical- (TO/LO) and acoustic- (TA) phonon replicas of the bulk free-exciton (FE) at \SI{1.098}{\eV} and \SI{1.137}{\eV}, respectively \cite{vouk_TwophononAssisted_1977}. In addition, the two-phonon (TO + O$^\mathrm{\Gamma}$) replica of the FE$_\mathrm{LO/TO}$ signal is weakly observed at \SI{1.034}{\eV} \cite{vouk_TwophononAssisted_1977}.

In sample A the Ge WL from the areas between the ordered dots contributes a no-phonon peak (WL$_\mathrm{NP}$) and the TO replica hereof (WL$_\mathrm{TO}$) at \SI{1.08}{\eV} and \SI{1.02}{\eV}, respectively \cite{brehm_QuantitativeDetermination_2008, brehm_KeyRole_2009, wachter_PhotoluminescenceHigh_1993}. No apparent WL signal is observed from sample B. Finally, rather broad QD signals appear in both samples between \SI{0.82}{\eV} and \SI{0.97}{\eV} with a NP peak at \SI{0.92}{\eV} and a TO replica (QD$_\mathrm{TO}$) at \SI{0.88}{\eV} \cite{tsybeskov_SiliconGermaniumNanostructures_2009, abstreiter_GrowthCharacterization_1996}. At \SI{10}{\kelvin} the QD-related signals of the two samples are very similar both in spectral appearance and with regard to the signal intensity. In agreement with our simulations, showing that with a thin spacer layer there is no localization of electrons between the QDs possible, this finding indicates that the lower QD in sample B does not contribute to the PL emission.

\subsection{Spatially resolved PL measurements}
\begin{center}
\begin{figure}
  \includegraphics[width=\linewidth]{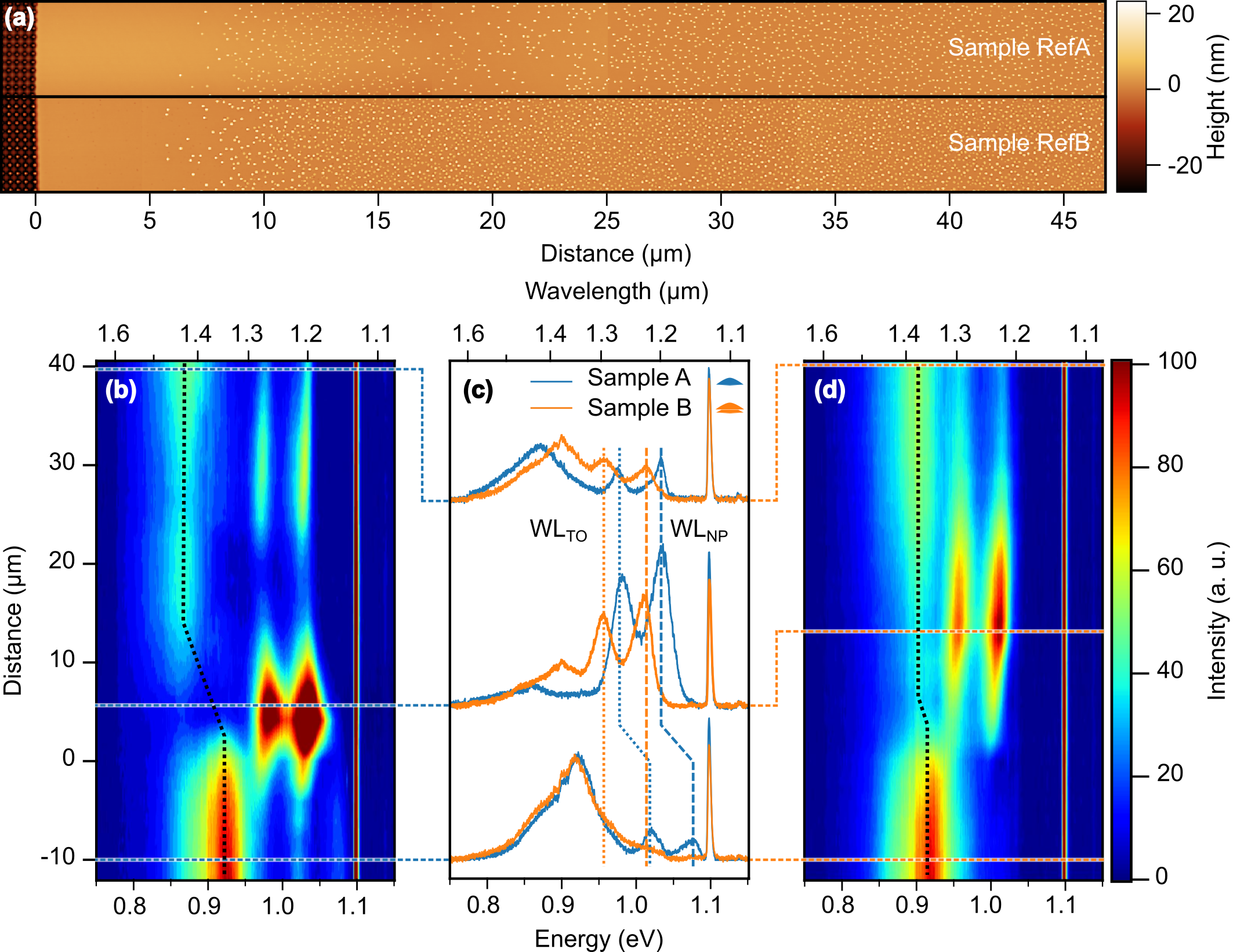}
  \caption{(a) AFM images of the uncapped samples RefA and RefB in the boundary regions between pit-patterned (\SI{340}{\nano\metre} period, \SI{100}{\nano\metre} radius) and plain substrate areas. The onset of the pit-pattern defines the origin of the distance scale, with positive values pointing into the plain substrate area. In both samples, a denuded zone free of QDs is observed in the first \SIrange{5}{7}{\micro\metre} of the plain area, followed by random QD nucleation with a bimodal shape distribution of pyramid- and dome-shaped QDs. (b), (d): Photoluminescence contour plots of sample A and B, respectively, on the same distance scale as in (a). As a guide to the eye, the dotted vertical lines indicate the respective shifts of the QD signals from the patterned to the plain areas. (c) Comparison of PL spectra of the two samples from the ordered field, the denuded zone and the randomly nucleated area (bottom to top), as indicated by the horizontal dashed lines in (b), (d). The dashed vertical lines indicate the energetic shifts of the respective WL-signal peaks. All spectra were recorded at \SI{10}{K}. For more details see the Suppl. Informations}
  \label{fig:contour}
\end{figure}
\end{center}

In Fig. \ref{fig:peaks} sample B lacks an apparent WL signal even though WLs associated with either of the QD layers are clearly resolved in the STEM and HAADF images (Fig. \ref{fig:TEM}). To gain further insight into this finding, we performed spatially resolved PL measurements in the near vicinity of ordered QD arrays, both on sample A and B (Fig. \ref{fig:contour}). Details of these experiments are discussed the Supplemental Information to this paper.

In brief, on the way from a pit-patterned region to the surrounding plain substrate area, deterministic QD ordering turns into random QD nucleation \cite{grydlik_UnrollingEvolution_2013}. For single-QD layers it is well known that the PL signals of both, the QDs and the WL, become red-shifted in the random-nucleation areas as compared to the pit-patterned regions \cite{grydlik_UnrollingEvolution_2013}. This behavior is attributed to enhanced alloying of QDs in a pit \cite{schmidt_MultipleLayers_2000, zhang_StrainEngineering_2010, grydlik_UnrollingEvolution_2013}, and pit-wall-decoration on the cost of the WL \cite{chen_InitialStage_2006}, respectively. 
For sample B we found a significantly smaller red-shift of the QD signal and only one WL signal in the plain substrate areas (Fig. \ref{fig:contour}(b)-(d)). From the observed shifts of the PL signals we could estimate that in sample A $\approx$2/3 of a Ge ML are transferred from the WL to the pit-walls. For sample B we found that only the upper WL contributes to the WL signal from the plain area, which we found to be about \SI{0.4}{ML} thicker than the lower one, presumably because of diffusion during the growth of the second QD layer. The absence of an apparent WL PL signal in the ordered regions of sample B is most likely caused by a more efficient transfer of photo-excited holes from the WL regions into the upper QD of the stack. 

\subsection{Excitation-intensity-dependent PL experiments}
We studied the dominating recombination mechanism of the photo-excited electron-hole pairs in samples A and B by conducting PL experiments as a function of the excitation intensity $I_\mathrm{E}$. Details of the experiment are given in the Supplementary Information. 

Over a large range of excitation intensities, we found in both samples a power-law behavior of the integrated QD signal $I_\mathrm{PL}$ of the form $I_\mathrm{PL} \propto I_\mathrm{E}^m$. We found for both samples identical exponents of $m = 0.7$. This value is typically observed in SiGe QDs, and has been associated with Auger recombination being the dominant recombination mechanism \cite{tsybeskov_SiliconGermaniumNanostructures_2009}.

\subsection{Temperature-dependent PL experiments}
To assess the temperature-stability of the PL signal, temperature-dependent \textmu-PL (T$_\mathrm{PL}$) measurements were performed in the same experimental run under identical measurement conditions for samples A, B, and C (Table \ref{tab:samples}). We chose an excitation intensity of \SI{14.1}{\kilo\watt\per\centi\metre\squared}, and the respective QD arrays with a period of \SI{340}{\nano\metre} and a pit radius of \SI{100}{\nano\metre} in all three samples.

\begin{figure}
\begin{center}
  \includegraphics[width=0.6\linewidth]{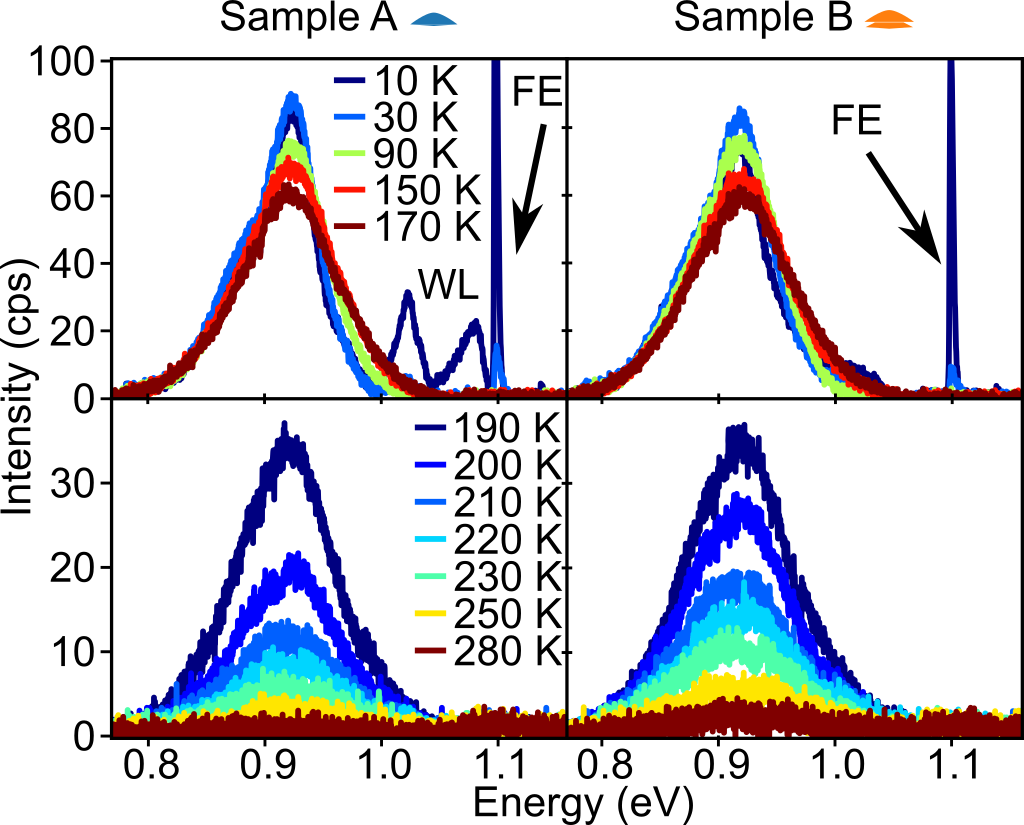}
	\caption{Temperature-dependent \textmu-PL measurements. Spectra shown in the left half of the figure are from the single-layer sample A, those in the right half from the stacked double-QD sample B with a Si spacer width of \SI{2}{\nano\metre}. The temperature range from \SIrange{10}{170}{\kelvin} is displayed in the upper panels, the range from \SIrange{190}{280}{\kelvin} in the respective lower panels. Note the different span of the intensity scales of the upper and lower panels. The FE$_\mathrm{LO/TO}$ signal from the Si substrate and the WL signal in sample A rapidly disappear with increasing temperature. While the intensities of the QD-related signal between \SI{0.8}{\eV} and \SI{1.0}{\eV} are comparable in both samples, the decay above \SI{190}{\kelvin} is significantly more pronounced in the single-QD-sample A.}
  \label{fig:tdep}
\end{center}
\end{figure}

Figure \ref{fig:tdep} shows the PL spectra of single-QD sample A (left panels) and on the right those of sample B with the \SI{2}{\nano\metre} thick spacer. The spectra in the temperature range from \SIrange{10}{170}{\kelvin} (respective upper panels) and from \SIrange{190}{280}{\kelvin} (lower panels) are plotted separately to account for the decrease of the PL signals with increasing T. The spectra of both samples are to scale in units of counts per second (cps). 

In the low-T range up to \SI{170}{\kelvin} (upper panels in Fig. \ref{fig:tdep}), the intensities of the QD-related signals between \SI{0.8}{\eV} and \SI{1.0}{\eV} are very similar in both samples. Above \SI{200}{\kelvin}, however, the signal decay is significantly more pronounced in sample A with its single-QD layer. At \SI{280}{\kelvin}, only the QD stack shows an evident, albeit weak, QD-related signal.

\begin{figure}
\begin{center}
  \includegraphics[width=0.6\linewidth]{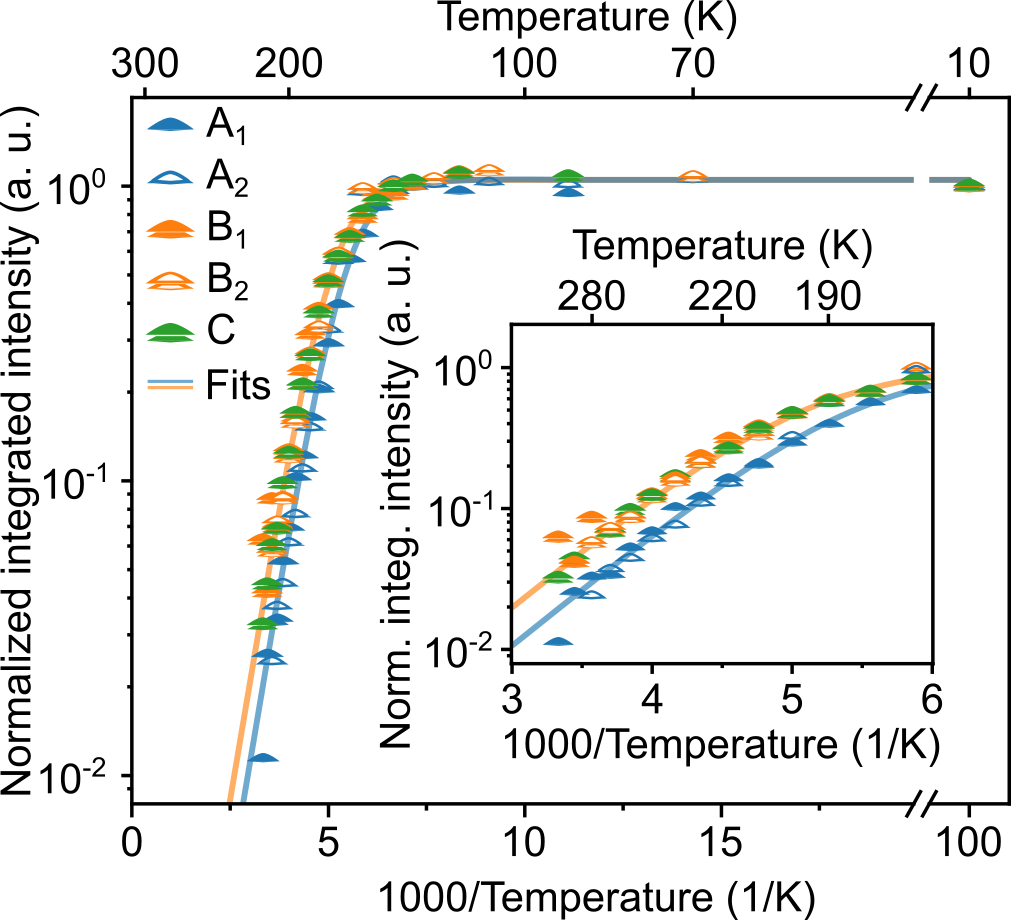}
  \caption{Arrhenius plot of the QD-related signals of two different measurements of the two samples from Fig. \ref{fig:tdep} and of sample C with a \SI{4}{\nano\metre} thick Si spacer. The PL signals are integrated between \SI{0.8}{\eV} and \SI{1.0}{\eV} and plotted over the investigated temperature range from \SIrange{10}{280}{\kelvin}. The inset is a zoom-in of the temperature range between \SI{170}{\kelvin} and \SI{300}{\kelvin}. Above \SI{220}{\kelvin} the integrated PL signals of the double-QD samples B and C exceeds the one of the single-QD-sample A by a factor of 2. The solid lines in the inset are fits to the two-level model discussed in the text.}
  \label{fig:arrhenius}
\end{center}
\end{figure}

To quantify the T-dependence of the QD-related PL signal, we integrated the PL intensity between \SI{0.8}{\eV} and \SI{1.0}{\eV} for each temperature and normalized it to the intensity at \SI{10}{\kelvin}. An Arrhenius plot is depicted in Fig. \ref{fig:arrhenius} for samples A, B and C, showing the normalized integrated QD signal on a logarithmic scale versus 1000/T. Sample A and B were measured a second time for proof of reproducibility (A$_1$, A$_2$ and B$_1$,B$_2$). The integrated intensities of all three samples increase slightly with temperature up to about \SI{170}{\kelvin} \cite{brehm_PhotoluminescenceInvestigation_2015, grydlik_OpticalProperties_2015}, and then drop rapidly towards higher temperatures. The range above \SI{200}{\kelvin} is expanded in the inset of Fig. \ref{fig:arrhenius} to highlight the differences of the three samples in this temperature range. Above approximately \SI{220}{\kelvin}, the integrated QD signals from the QD stacks (samples B and C) are about a factor of two more intense than the signal from the single-layer-sample A. In this temperature range, the integrated intensities decay with very similar, linear slopes in the Arrhenius plot.

The solid lines in Fig. \ref{fig:arrhenius} are fits to the data points using the simple two-state model \cite{wachter_PhotoluminescenceHigh_1993} (see "Methods" section).

\begin{equation}
	I(T) = I_0\cdot(1+A\cdot \exp(-E_\mathrm{A}/k_\mathrm{B}T))^{-1}
\label{equ:Arr}
\end{equation} 

The low-temperature limit $I(T\rightarrow0)= I_0$ was taken at \SI{10}{\kelvin}, the lowest temperature measured. We optimized the fits in the relevant temperature range above \SI{190}{\kelvin}, where thermally activated behavior is clearly observed. Under these conditions, the fits depicted in Fig. \ref{fig:arrhenius} correspond to ($E_A$ = \SI{158}{\milli\eV}; $A$ = 23000) for sample A, and ($E_A$ = \SI{158}{\milli\eV}; $A$ = 12000) for sample B and C. For the fits, both measurement series were taken into account. While the activation energies for thermal depopulation of the QD states are essentially identical, the fitted pre-factors $A$ = $\frac{g_2\tau_1}{g_1\tau_2}$ (equ. \ref{equ:A} in the "Methods" section) differ by a factor of about 2. The ratio $g_2/g_1$ of the densities of states should be very similar for the three samples, as they describe in all three cases the electron-hole pair bound to a SiGe QD ($g_1$) and the dissociated electron-hole pair with the electron delocalized in the conduction band of the surrounding Si matrix ($g_2$). Therefore, the significant differences of the experimentally determined $A$ reflect the different lifetime ratios $\tau_1/\tau_2$ for the two samples. $\tau_2$ is expected to describe essentially the lifetime of the delocalized electrons \cite{wachter_PhotoluminescenceHigh_1993} which is expected to be comparable in all three samples. Hence, the lifetime $\tau_1$ of the electron-hole pair located at the QD is the most important contribution to the respective values of $A$. The different values of $A$ extracted from the experiments imply that the lifetime of the bound QD states in the double-QD samples B and C is about a factor of 2 shorter than in sample A. This is a strong indication of a significantly enhanced electron-hole interaction and an increased PL yield for the closely stacked QD pairs as compared to the single QDs in the reference sample.

\section{Summary and Concluding Remarks}

In summary, we addressed the problem of inefficient optical recombination in the indirect-gap, type-II Si/SiGe heterostructure by the investigation of laterally ordered, closely stacked SiGe QD pairs. In this approach, we aimed at a functional separation of the role of the two QDs in a stack: The role of the lower QD is that of a pure stressor, whereas only the upper QD facilitates radiative exciton recombination. The stressor-QD creates a preferential nucleation site for the upper one, but its main purpose is controlled straining of both, the upper QD and the adjacent Si cap. In this way, the hole wave function can be shifted toward the apex of the upper QD, and the electrons become stronger localized in the strain pocket in the Si cap above the upper QD. As a result of this particular strain engineering approach, enhanced wave function overlap can be achieved, concomitant with a stronger electron-hole interaction and increased PL yield. Indeed, we find that the double QD arrangement shifts the thermal quenching of the photoluminescence signal to significantly higher temperatures. Moreover, the closely stacked QD-pair efficiently suppresses the detrimental radiative recombination channel provided by the Ge WL in the areas between the ordered QDs. As our approach with spacer-layer thicknesses smaller than \SI{5}{\nano\metre} deactivates optical recombination at the lower QD by preventing electrons to become localized between the QDs, the QD stack remains in principle suitable as a single-photon source that can be positioned with nm precision in a CMOS-compatible Si device environment. 

With these promising results, a next step to further enhance the PL efficiency of this approach will be further optimization of strain tuning. For example, more strain, and thus stronger electron confinement, could be achieved by conserving the shape of the lower QD by using lower growth temperatures for the upper QD layer. Furthermore, the implantation with Ge-ions to create deep-level defects in the upper QD of the stack \cite{grydlik_LaserLevel_2016} will combine perfect lateral ordering with strain tuning of the upper QD and a quasi-direct optical transition via the implanted defect \cite{murphy-armando_LightEmission_2021}, which will be addressed in future works.

\section*{\Huge Methods}

\section*{Sample Fabrication}

In the first step, \SI[product-units = single]{17.5 x 17.5}{\milli\metre\squared} samples were cut from a 4" high-resistivity (R$_0>$\SI{1000}{\ohm\centi\metre}) float-zone (FZ) Si$\langle$100$\rangle$ wafer. After chemical cleaning in organic solvents and oxide removal in \SI{1}{\percent} fluoric acid (HF), the samples were spin-coated with a positive polymethylmethacrylat (PMMA) resist (Allresist 679.04). The pit-pattern for the preferential nucleation of the SiGe QDs was defined by means of electron beam lithography using a Raith eLine system operated at an acceleration voltage of \SI{30}{\kilo\volt} and a beam current of \SI{300}{\pico\ampere}. After developing in Allresist 600-56, the remaining resist served as a mask for reactive ion etching in an Oxford Plasmalab 100 ICP system using a gas flow of \SI{2}{sccm} for SF$_6$ and \SI{1}{sccm} for O$_2$ at a cryogenic sample temperature of \SI{-90}{\degreeCelsius}. A large number of pit arrays were realized simultaneously on each sample. Hereby, the pit radii were varied in six steps from \SIrange{75}{100}{\nano\metre}, and the pit periods in 13 steps from \SIrange{340}{490}{\nano\metre} (pit densities ranging from \SIrange{4.16e8}{8.65e8}{\per\centi\metre\squared}) from array to array. Each pit field contained a square lattice of \SI[product-units = single]{80 x 80} pits etched to a uniform depth of \SI{50}{\nano\metre} to allow for single QD nucleation in each pit during subsequent MBE growth \cite{schatzl_EnhancedTelecom_2017, brehm_SitecontrolledAdvanced_2017}.

After removal of the remaining resist in an oxygen plasma followed by piranha etching \cite{caro_ZurKenntniss_1898}, an RCA cleaning step \cite{kern_CleaningSolution_1970}, and a final dip in \SI{5}{\percent} HF, the pre-patterned substrates were introduced via a load-lock system into our Riber Siva 45 MBE system with electron beam evaporators for Si and Ge \cite{brehm_PhotoluminescenceInvestigation_2015}. After a degassing step for \SI{20}{\minute} at \SI{650}{\celsius}, a \SI{50}{\nano\metre} thick Si buffer layer was grown at a rate of \SI{0.7}{\angstrom\per\second} during a ramp-up of the substrate temperature from \SIrange{470}{570}{\celsius}. Under these conditions, the Si buffer converts the cylindrically shaped pits into inverted pyramids with \{1~1~10\} facetted sidewalls \cite{chen_InitialStage_2006}. The facetted pits act as nucleation sites for the subsequent growth of SiGe QDs \cite{zhong_TwodimensionalPeriodic_2003, brehm_PhotoluminescenceInvestigation_2015, brehm_SitecontrolledAdvanced_2017, karmous_GeDot_2004}. For the first QD layer, the temperature was increased to \SI{670}{\celsius}, and \SI{5.6}{\angstrom} of pure Ge were deposited at a rate of \SI{0.04}{\angstrom\per\second}. For sample A (see Table \ref{tab:samples}), a \SI{70}{\nano\metre} thick silicon cap layer was then grown at \SI{470}{\celsius} and \SI{0.12}{\angstrom\per\second}. For samples B–D, a silicon spacer layer was grown at \SI{470}{\celsius} after the first QD layer, with its thickness being varied between \SI{2}{\nano\metre} and \SI{5}{\nano\metre} (Table \ref{tab:samples}). This rather low growth temperature was chosen to preserve conformal covering of the first QD layer \cite{brehm_InfluenceSi_2011}. Subsequently, the second QD layer was grown by depositing \SI{4.2}{\angstrom} of Ge at the otherwise same conditions as the first one. Due to the built-in strain, the QDs of the second layer form directly on top of the under-lying QDs, thus becoming vertically stacked \cite{schmidt_MultipleLayers_2000, zhang_StrainEngineering_2010}. Finally, the QD stacks were capped by \SI{50}{\nano\metre} of Si deposited at a rate of \SI{0.7}{\angstrom\per\second} during a temperature ramp from \SIrange{470}{500}{\celsius}. 

\section*{Scanning Transmission Electron Microscopy}

Scanning transmission electron microscopy (STEM) was performed in a JEOL JEM-2200FS instrument at \SI{200}{\kilo\electronvolt} primary beam energy using either a bright field (BF) or a high angle annular dark field (HAADF) detector. Sample preparation for STEM was conducted by focused ion beam (FIB) in a Zeiss 1540 XB instrument. Before lamella cutting with \SI{30}{\kilo\electronvolt} Ga-beam energy, an electron-beam-stimulated Pt-deposition was used to protect the sample surface. The final thinning was performed at an energy of \SI{5}{\kilo\electronvolt} to reduce specimen amorphization. Lamellae were prepared from sample D (Table 1) from both the pre-patterned and the plain areas of the sample, as well as from sample RefB. In the pre-patterned areas lamellae were cut along the \textlangle110\textrangle-orientation of the pit rows, as well as in \textlangle100\textrangle-direction, i.e., under \ang{45} with respect to the pit-row orientation.

\section*{Nanotomography}

For a comprehensive assessment of the composition profiles in the QD-stacks, we performed nanotomography \cite{rastelli_ThreeDimensionalComposition_2008} via a combination of repeated wet chemical etching steps and AFM scans. The samples were immersed into a NHH solution (1:1 vol. (\SI{28}{\%} NH$_\mathrm{4}$OH):(\SI{31}{\%} H$_\mathrm{2}$O$_\mathrm{2}$)), which exhibits an exponentially increasing etching rate with increasing Ge-content \cite{rastelli_ThreeDimensionalComposition_2008}. Expecting a higher Ge-fraction in the upper parts of the QDs, the etching times were increased for each step from \SIrange{1}{1015}{min} to achieve approximately equidistantly spaced profiles from the AFM scans recorded after each etching step.

The etching profiles were converted into maps of the Ge content using the methods described in Ref. \cite{rastelli_ThreeDimensionalComposition_2008}. For image correction, overlay, tomography computation and visualization, a software package provided by A. Rastelli was employed. 

\section*{Micro-Photoluminescence Measurements}

The samples were studied by micro-photoluminescence (\textmu-PL) spectroscopy in a setup described in detail elsewhere \cite{spindlberger_AdvancedHydrogenation_2021}. For excitation a continuous-wave (cw) diode laser with a wavelength of \SI{441}{\nano\metre} (\SI{532}{nm} frequency-doubled Nd:YAG laser for excitation-intensity-dependent measurements) was focused via a microscope objective with a numerical aperture of 0.7 onto a spot on the sample with a diameter of approximately \SI{3}{\micro\metre} and an adjustable excitation-intensity of up to \SI{1100}{\kilo\watt\per\centi\metre\squared}. Collected by the same objective, the \textmu-PL signal was guided through a multimode fiber to a \SI{300}{\milli\metre} grating spectrometer (\SI{300}{grooves/\milli\metre}) equipped with a liquid-nitrogen-cooled InGaAs line-detector from Princeton Instruments. For comparative experiments, several samples were mounted simultaneously on the cold-finger of a continuous-flow helium cryostat that allowed adjustment of the sample temperatures between \SI{10}{\kelvin} and \SI{300}{\kelvin}. In this way, temperature- and excitation-intensity-dependent measurements could be performed under identical excitation and imaging conditions on samples B and C with stacked QDs and on the single-layer-sample A.

\section*{nextnano++ Simulations}

The commercial nextnano++ software package \cite{birner_NextnanoWww_} is essentially a Schr\"odinger/Poisson solver with built-in data base for various heteromaterials. nextnano++ provides a comprehensive means for assessing the optical and electronic properties of strained semiconductor heterostructures of almost arbitrary composition. In a first step, a tentative, three-dimensional (3D) composition model of the desired heterostructure is defined as an input file for the simulation. The program then provides the corresponding strain distribution under the assumption of coherent growth, i.e., in the absence of plastic strain relaxation. The composition and strain topologies determine the 3D variation of the band structure in the simulation volume. Finally, Schr\"odinger/Poisson iterations in a single-band effective mass approach yield the electronic and optical properties of the initially defined heterostructure. Comparison with experiments then leads to modifications of the initial composition model as an input for further iterations.

For reasons of simplicity, the \{1~0~5\}-facetted pits \cite{chen_InitialStage_2006} as well as the multi-facetted, dome-shaped QDs \cite{medeiros-ribeiro_ShapeTransition_1998} were treated as combinations of rotational symmetric cones and truncated cones. To tailor the Ge-profile according to the nanotomography results, an sequence of 5 cones with increasing size and Ge-content were superimposed for the lower and the upper part of the dots, respectively, each with an linear Ge-gradient in growth direction. The lower part of both, single layer and upper dot of the stack, consists of truncated cones with an inclination angle of \ang{25.3}, corresponding to a \{1~1~3\} facet, while the top of the QDs was simulated by another set of cones with a geometry corresponding to a \{1~0~5\} facet with an inclination angle of (\ang{11.3}). The grid for the Schr\"odinger/Poisson solver was adapted to the geometry to facilitate a compromise between resolution and computing time. Emphasis was put on details near the apices of the QDs.

Since the nanotomography experiments had to be conducted on the un-capped reference samples RefA and RefB, the Ge content on the surface of the upper QD is expected to be overestimated by the segregation of about two monolayers of Ge \cite{schmidt_MultipleLayers_2000}. As a significant part of the segregated Ge will diffuse into the cap upon overgrowth with Si, we corrected the Ge content on the surface of the nanotomography maps by iteratively comparing the simulated optical transition energies with the \textmu-PL results. For this purpose, a starting value for the maximum Ge concentration was chosen, and the energy for the lowest transition was calculated via nextnano++. This value was compared to the QD-related PL signal which has its maximum in the ordered region of sample B at \SI{0.92}{\electronvolt} (Fig. \ref{fig:peaks}). We achieved convergence of the iteration with a maximum Ge content of approximately 33\% at the apices of the QDs.

\section*{Arrhenius fits}

To quantify the temperature-dependent behavior of the three samples, we fitted the data points with a simple two-state model with a single activation energy \cite{wachter_PhotoluminescenceHigh_1993, wan_EffectsInterdiffusion_2001}:

\begin{equation}
	I(T) = I_0\cdot(1+A\cdot \exp(-E_\mathrm{A}/k_\mathrm{B}T))^{-1}
\label{equ:Arr}
\end{equation} 

$I$($T$) is the integrated PL intensity as a function of the thermal energy $k_BT$; $I_0$ is the integrated PL intensity at $T\rightarrow0$, and $k_B$ is the Boltzmann constant. For $A$ $\gg$ 1, $I_0/A$ is the extrapolated PL intensity in the limit $1/T\rightarrow 0$. $E_A$ is the activation energy from the electron-hole pair bound at the QD (State 1) to a deactivated State 2, where the electron has become delocalized in the surrounding Si matrix \cite{wachter_PhotoluminescenceHigh_1993}. The pre-factor $A$ in the denominator reads
\begin{equation}
	A = \frac{g_2\tau_1}{g_1\tau_2}
\label{equ:A}
\end{equation}
where the $g_i$ are the densities of states of the two involved states, and the $\tau_i$ are the respective excited state lifetimes \cite{wachter_PhotoluminescenceHigh_1993}. If the dimensionalities of initial and deactivated final states are different, the ratio $g_2/g_1$ assumes an explicit temperature-dependence of the form \cite{wachter_PhotoluminescenceHigh_1993, grydlik_OpticalProperties_2015}:
\begin{equation}
	\frac{g_2}{g_1} = T^n.
\label{equ:g2/g1}
\end{equation}
In our configuration, we expect thermal deactivation from a zero-dimensional (0D) confined state at a QD into a 3D band state, which would be associated with $n$ = 3/2 \cite{wachter_PhotoluminescenceHigh_1993}. We found, however, that in the accessible temperature range a temperature-dependence according to equ. \ref{equ:g2/g1} has little influence on the quality of the fits. Therefore, we set $n$ = 0 in the following, i.e., we treated $A$ as a temperature-independent constant representing the densities of states and the lifetimes of the two states according to equ. \ref{equ:A}. 

\bibliography{Lib.bib}

\section*{Acknowledgements}
We thank A. Halilovic for help during sample processesing, A. Rastelli for providing the software for nanotomography and assistance in evaluation of the respective experiments. We further thank G. Hesser (ZONA) for STEM specimen preparation and T. Poempool for fruitful discussion.

This study was supported by the Austrian Science Fund (FWF) via projects no. P30564NBL, co-funded by the Province of Upper Austria and Y1238-N36 as well as the Linz Institute of Technology (LIT): Grant No. LIT-2019-7-SEE-114. Also, the financial support by the Austrian Federal Ministry for Digital and Economic Affairs, the National Foundation for Research, Technology, and Development, and the Christian Doppler Research Association is gratefully acknowledged.

\section*{Author contributions}
J.S. processed the samples and carried out the experiments and simulations. J.A., L.V. and M.B. conducted MBE growth. H.G. carried out STEM and HAADF characterizations. L.S. contributed to PL experiments. The manuscript was written by J.S. and F.S. with contributions of all authors.
 
\section*{Competing interests}
The authors declare no competing interests.

\end{document}